\documentstyle[amsfonts,multicol,fancyheadings,eqsecnum,prd,aps]{revtex}

\def\buildchar#1#2#3{\null \! \mathop {\vphantom 
{#1}\smash #1}\limits ^{#2}_{#3}\!\null }
\def\OT#1{\buildchar{{#1}}{\;_\sim}{}\/} 
\def\UT#1{\buildchar{{#1}}{}{^\sim}\/}
\def\OTT#1{\buildchar{{#1}}{\;_\approx}{}\/} 
\def\UTT#1{\buildchar{{#1}}{}{^\approx}\/} 

\begin{document}
\draft

\title{The gauge group in the real triad formulation of general relativity}

\author{J.\ M.\ Pons
\footnote[1]{Electronic address: pons@ecm.ub.es}} 
\address{Departament d'Estructura i Constituents 
de la Mat\`eria, Universitat de Barcelona,\\
and Institut de F\'\i sica d'Altes Energies,\\ 
Diagonal 647, E-08028 Barcelona, Catalonia, Spain} 

\author{D. C. Salisbury
\footnote[2]{Electronic address: dsalisbury@austinc.edu}} 
\address{Department of Physics,
Austin College, Sherman, Texas 75090-4440, USA} 

\author{L.\ C.\ Shepley
\footnote[3]{Electronic address: larry@helmholtz.ph.utexas.edu}} 
\address{Center for Relativity, Physics Department, \\ 
The University of Texas, Austin, Texas 78712-1081, USA \\ 
~} 

\date{Version of 22 December 1999\ --- to be submitted to 
\textit{Gen.\ Rel.\ Grav.}}
\maketitle

\begin{abstract}
We construct explicitly generators of projectable four-dimensional
diffeomorphisms and triad rotation gauge symmetries in a model of
vacuum gravity where the fundamental dynamical variables in a Palatini
formulation are taken to be a lapse, shift, densitized triad,
extrinsic curvature, and the time-like components of the Ricci
rotation coefficient.  Time-foliation-altering diffeomorphisms are not
by themselves projectable under the Legendre transformations.  They
must be accompanied by a metric- and triad-dependent triad rotation. 
The phase space on which these generators act includes all of the
gauge variables of the model.

\end{abstract}

\pacs{04.20.Fy, 04.40.-b, 11.10.Ef, 11.15.-q \hfill gr-qc/9912087} 

\setlength{\columnseprule}{0pt}\begin{multicols}{2}


\section{Introduction}
\label{sec:intro}

General covariance is the fundamental symmetry of the classical
Lagrangian formulation of general relativity.  One might be tempted to
think that it is destroyed in the Hamiltonian version of the theory in
which, ostensibly, time is distinguished from space.  In recent works
we have established a general framework for analyzing and describing
the preservation of local (gauge) symmetries under the Legendre map
from configuration-velocity space (the tangent bundle) to phase space
(the cotangent bundle)\cite{pons/salisbury/shepley/97}.  The program
was applied to general relativity using conventional metric variables
and to Einstein-Yang-Mills theory\cite{pons/salisbury/shepley/99a}. 
In the former case it was shown that the four-dimensional
diffeomorphism group is not preserved under this map.  Rather,
infinitesimal elements of the gauge group contain a compulsory
dependence on the lapse and shift functions.  In the
Einstein-Yang-Mills case gauge transformations also involve internal
gauge transformations.  Nevertheless there is a sense in which all
diffeomorphisms act as transformations on the full phase space of the
theory, since any such transformation may be realized with some metric
plus Yang-Mills field.

In this paper we apply the program to a tetrad version of Einstein's
general theory of relativity.  Spinorial and tetrad formulations of
general relativity were introduced initially with the intention of
coupling fermionic matter fields to gravity in an eventual quantum
theory of gravity.  The emerging Dirac-Bergmann constraint algorithm
was applied to Lagrangian or Hamiltonian models by Heller and Bergmann
\cite{heller/bergmann/51}, DeWitt and DeWitt\cite{dewitt/dewitt/52},
Dirac \cite{dirac/62}, and Schwinger \cite{schwinger/63}.  In these
early investigations an effort was not made to retain the local
Lorentz freedom to rotate and boost the tetrad axes.  Schwinger chose
one of the vectors of the tetrad to point perpendicular to the
equal-time hypersurfaces.  The coordinate time was also taken over in
the Hamiltonian model as the evolutionary time.  Consequently, all
that remained were local triad rotations tangential to the equal-time
hypersurfaces.  This will actually be our point of departure below. 
Later, several authors have been concerned with retaining the full
local Lorentz group\cite{nelson/teitelboim/78,%
henneaux/83,charap/nelson/86,charap/henneaux/nelson/88,%
henneaux/nelson/schomblond/89,maluf/91,lusanna/russo/98}.  Generally,
retention of the local Lorentz group is achieved by adding auxiliary
pure gauge boost variables.  We will not pursue this direction here,
although our analysis can easily be generalized to include the full
local Lorentz symmetry.  Clayton analyzed the symmetry of triad models
only under spatial diffeomorphisms \cite{clayton/98}.

Interest in time-foliation-conforming triads has surged recently. 
This interest stems from Ashtekar's approach to general relativity
\cite{ashtekar/86,ashtekar/87,ashtekar/91}.  The relation of Ashtekar
variables to triads was elucidated by Goldberg \cite{goldberg/88},
Friedman and Jack \cite{friedman/jack/88}, and Henneaux, Nelson, and
Schomblond \cite{henneaux/nelson/schomblond/89}.

Our novel contribution in this paper to this extensive literature has
to do with the gauge group stemming from the full four-dimensional
diffeomorphism group with local triad rotations appended where
necessary.  We show that this group is retained under the Legendre map
to phase space.  This group acts as a transformation group on all of
the dynamical variables of the theory (including pure gauge
variables).  In particular, time evolution is recognizable as a
symmetry transformation on members of equivalence classes (under the
full four-dimensional diffeomorphism group) of solutions of the
dynamical equations.  The implications for an eventual quantum theory
of gravity and its associated problem of time, we feel, are profound:
Any reasonable quantization procedure, whether it take as fundamental
gravitational variables the metric, tetrad, or triad fields, must
respect and presumably exploit this symmetry.

Our plan is as follows.  Our focus is local; the analysis is strictly
applicable to spatially compact spacetimes.  We work within a 3+1
spacetime foliation; the fundamental fields in this model include
triads, lapse functions, and shift vectors.  Our dynamical treatment
also includes some of the time-like terms of the Ricci rotation
coefficients as independent fields in a Palatini-like variational
approach.  We show that the full diffeomorphism-induced gauge group is
present, and we display explicitly the corresponding generators on the
full phase space (which includes gauge variables).

After completing the Lagrangian and Hamiltonian descriptions in
Section \ref{sec:action}, we review the general projectability
requirements in Section \ref{sec:project}, deducing that coordinate
transformations for which $\delta x^{0} \neq 0$ are not by themselves
projectable.  A gauge rotation constructed with the aid of the
four-dimensional Ricci rotation coefficients must be appended to them. 
Next, in Section \ref{sec:generators} we determine the variations
engendered by the secondary constraints.  We demonstrate that it is
possible to determine the structure coefficients of the gauge group
algebra from knowledge of the action of the gauge group on
configuration-velocity variables alone, without reference to Poisson
brackets.  Finally in Section \ref{sec:symgens} we display the full
set of gauge-symmetry generators.  In Section \ref{sec:conclusions} we
summarize and discuss implications of the work and future extensions,
such as to the Ashtekar formulation
itself\cite{pons/salisbury/shepley/99b}.

\section{The tetrad action}
\label{sec:action}

We take as our fundamental dynamical variables the tetrad one form
fields $e^{I}_{\mu}$, with inverse $E^{\mu}_{I}$, yielding the metric
$g_{\mu \nu}=e^{I}_{\mu}e_{I\nu}$.  The greek coordinate indices range
from 0 to 3, as do the upper case latin orthonormal basis labels.  The
Minkowski index labels are raised and lowered with the Minkowski
metric $\eta^{IJ}=\eta_{IJ}$, which we take to have signature
$-\,$$+\,$$+\,$$+$.  Lower case latin indices from the beginning of
the alphabet will represent spatial coordinates, while lower case
latin indices from the middle of the alphabet will denote orthonormal
triad labels (and therefore may be raised or lowered using
$\delta_{ij}$).

The Ricci rotation coefficients $\Omega_{\mu}^{IJ}$ provide a
Lorentz-group connection which leaves the tetrad covariantly constant
and are given by
\begin{equation}
\Omega_\mu^{IJ} = E^{\beta I} e^J_{\beta ,\mu} 
- \Gamma^\beta_{\gamma \mu} E^{\gamma I}e^J_\beta \ , 
\label{ricrot}
\end{equation}
where the $\Gamma^\beta_{\gamma \mu}$ are the Christoffel symbols. 
The associated curvature is
\[
{}^{4}\!R_{\mu \nu}^{IJ} = 2 \partial_{[\mu}\Omega_{\nu]}^{IJ} +2
\Omega_{[\mu}^{IM} \Omega_{\nu]M}^{J} \ .  \] A convenient action
expressed in terms of these variables is one half the Hilbert action,
\begin{equation}
{1 \over 2}\int d^{4}x \, \sqrt{|^{4}g|}\, {}^{4}\!R_{\mu \nu}^{IJ} 
E^{\mu}_{I}E^{\nu}_{J}\ .
\end{equation}
We make a $3+1$ decomposition of the tetrad in terms of triad
$T^{a}_{i}$, lapse $N$, and shift $N^{a}$ as follows: \[ E^\mu{}_I =
\left(\begin{array}{cc} N^{-1} &0\\
-N^{-1} N^a & T^a_i
\end{array} \right)\ ,
\]
with inverse
\[
e^I{}_\mu
= \left(\begin{array}{cc}
N &0\\
t^i{}_a N^a & t^i{}_a
\end{array} \right)\ ,
\]
where $t^{i}_{a}T^{b}_{i}=\delta^{a}_{b},\
t^{i}_{a}T^{a}_{j}=\delta^{i}_{j}$.

In a succeeding paper, we will be applying our technique to the
Ashtekar formalism in full \cite{pons/salisbury/shepley/99b}, and so
at this point we adopt notation with that end in mind.  We introduce
as some of our fundamental dynamical configuration variables the
components of a contravariant triad field with density one under
spatial diffeomorphisms:
\[
\OT T^{a}_{i} := t T^{a}_{i}\ ,
\]
where $t$ is the determinant of $t^{i}_{a}$.  We also will represent
densities of arbitrary positive or negative weight with appropriate
numbers of over- or under-tildes, respectively.  We also note that \[
\Omega_{a}^{0i} = T^{bi} K_{ba} =: K^{i}_{a}\ , \] where $K_{ab}$ is
the extrinsic curvature.

The Lagrangian density expressed in terms of these new variables,
after subtracting a total spatial divergence, is (where $\dot{}$ is
partial with respect to time)
\begin{eqnarray}
{\cal L} &=& - K^i_a\dot{\OT T}{}^a_i
+ \epsilon^{ijk} \OT T^a_i K^j_a \Omega^k_{0} -2 N^a 
\OT T^b_i D_{[a} K^i_{b]}
			 \nonumber \\
&& +{1\over2} \UT N \OT T^a_i \OT T^b_j
({}^{3}\!R^{ij}_{ab} + K^i_a K^j_b - K^i_b K^j_a) \ . 		 
\label{lagrangian}
\end{eqnarray}
We translate from one triad label index to two index antisymmetric
objects by taking the dual using the three-dimensional Levi-Civita
symbol $\epsilon^{ijk}$; for example, \[ \Omega^{i}_{0}=
{1\over2}\epsilon^{ijk}\Omega^{jk}_{0}\ .  \] In the relation above
$D_a$ denotes the covariant derivative formed with the
three-dimensional rotation coefficients and Christoffel symbols, so we
have, for example,
\[
D_{a} T^{b}_{i}
=\partial_{a} T^{b}_{i}
+ {}^{3}\Gamma_{ca}^{b} T^{c}_{i}
+\omega_{a}^{ij} T^{b}_{j} = 0 \ ,
\]
where the 3-dimensional rotation coefficients \begin{eqnarray*}
\Omega_{a}^{ij} &=& \UTT g_{ac} \OT T^{b[i}\OT T^{j] c}_{~~,b} 
+ \OT T^{b[i} \UT t^{j]}_{c} \UT t^k_a \OT T ^c_{k,b} 
\\
&& +\UT t^{[i}_b \OT T^{j] b}_{~~,a}
+\UT t^k_c \OT t^c_{k,b} \UT t^{[i}_a \OT T^{j]b} 
\\
&=&: \omega_{a}^{ij}
\end{eqnarray*}
are constructed from the triad fields.

We will undertake a Palatini-type variation in which we vary $ \OT
T^{a}_{i}$, $K^{i}_{a}$, $N^{a}$, $\Omega_{0}^{i}$, and $\UT N$
independently.  This choice of variables is justified because the
variables $K^{i}_{a}$ and $\Omega_{0}^{i}$ can be considered as
independent auxiliary variables for the Lagrangian (\ref{lagrangian}):
That is, their equations of motion allow their determination in terms
of the other variables.

It may seem strange initially that we have identified
$\Omega_{0}^{ij}$ as an independent variable.  Examination of its
triad dependence in (\ref{ricrot}) reveals the rationale:
\begin{eqnarray} \Omega^{ij}_0 &=& \epsilon^{ijk}\Omega^{k}_{0} = T^{a
[i} \dot t^{j]}_{a} -g_{a\mu}\Gamma^{\mu}_{b0}T^{b [i} T^{j]a}
\nonumber \\
&=&\OT T^{a [i} \UT{\dot t}^{j]}_{a}
- g_{a\mu}\Gamma^{\mu}_{b0} T^{b [i} T^{j]a} 
\nonumber \\
&=& \OT T^{a [i} \UT{\dot t}{}^{j]}_{a}
- N^a_{,b} t^{[i}_a T^{j]b}
+ N^c t^k_c T^{a[i} T^{j]b} t^k_{a,b}
\nonumber \\
&&\qquad +N^c t^{[i}_{c,b} T^{j]b} \ .
\label{Om0ij}
\end{eqnarray}
The first term on the right hand side of (\ref{Om0ij}) can be varied
arbitrarily by adding SO(3) rotations to the time derivative of the
triad, so $\Omega^{ij}_{0}$ in fact can be taken to be an independent
gauge variable.  We shall see below, in (\ref{EoM}), that
$\Omega_{0}^{ij}$ is precisely this arbitrary rotation.  The
3-dimensional curvature ${}^{3}\!R^{ij}_{ab}$ is constructed from the
three-dimensional rotation coefficients $\omega_{a}^{ij}$.

Variation of the action yields the primary constraints in phase space
\begin{equation}
0 = P^{i}_{a}+K^{i}_{a}
= \OT \pi^{a}_{i}
= \OTT P
= \OT P_{a}
= \OT P_{i} \ ,
\label{prim}
\end{equation}
where $P^{i}_{a}$, $\OT \pi^{a}_{i}$, $\OTT P$, $\OT P_{a}$, and $\OT
P_{i}$ are the momenta conjugate to $\OT T^{a}_{i}$, $K^{i}_{a}$, $\UT
N$, $N^{a}$, and $\Omega^{i}_{0}$, respectively.

The first two constraints in (\ref{prim}) are second class in the
sense of Dirac \cite{dirac/50,dirac/64}; the variables $P^{i}_{a}$ and
$\OT \pi^{a}_{i}$ are eliminated in phase space, so that the Dirac
bracket becomes
\[
\{\OT T^a_i, K'^j_b \}
= -\delta^3 (x-x') \delta^a_b \delta^j_i \ , \]
leading to the canonical Hamiltonian density \begin{eqnarray*}
\OT{\cal H}_{ c}
&=& -\epsilon^{ijk} \OT T^a_i K^j_a \Omega^k_{0} 
+2 N^a \OT T^b_i D_{[a} K^i_{b]}
\\
&&-{1\over2} \UT N \OT T^a_i \OT T^b_j
({}^{3}\!R^{ij}_{ab} + K^i_a K^j_b - K^i_b K^j_a) \ . \end{eqnarray*}

Preservation in time of the remaining primary constraints leads to the
secondary constraints
\begin{eqnarray}
\OT{\cal H}_{i} &=& -\epsilon^{ijk} K^j_a \OT T^a_k =0 \ , 
\label{secdry.i} \\
\OT{\cal H}_{a} &=& -2 \OT T^b_i D_{[b} K^i_{a]} =0 \ , 
\label{secdry.a} \\
\OTT{\cal H}_{0} &=& -{1\over2} \OT T^a_i \OT T^b_j 
({}^{3}\!R^{ij}_{ab} + K^i_a K^j_b - K^i_b K^j_a) =0 \ . 
\label{secdry.0}
\end{eqnarray}
There are no more constraints.  Note that (\ref{secdry.i}) is just the
condition that $K_{ab}$ is symmetric.  Furthermore, (\ref{secdry.a})
can be rewritten in the familiar form
\[
D_a K^b_b - D_b K^b_a = 0 \ .
\]

We now turn to the canonical equations of motion.  The variation with
respect to $K^{i}_{a}$ of $ \int d^{3}x \, \OT{\cal H}_{c}$ is
straightforward, leading to:
\begin{eqnarray}
\dot{\OT T}{}^a_i
&=& -{\delta\over\delta K^i_a }\int d^3 x' \,\OT{\cal H}'_{c} 
\nonumber \\
&=& -\epsilon^{ijk} \OT T^a_j \Omega^k_{0} 
+2 D_b N^{[b} \OT T^{a]}_i
\nonumber \\
&&\quad + \UT N \OT T^a_i \OT T^b_j K^j_b 
- \UT N \OT T^b_i \OT T^a_j K^j_b \ .
\label{EoM}
\end{eqnarray}

The variation with respect to $\OT T^a_i$ requires more work.  We note
that the variation of $\omega^{ij}_a$ must be a tensor, so we can
replace ordinary derivatives by covariant derivatives.  Also, using
the fact that the covariant derivative of the triad is zero, we find
that
\begin{eqnarray*}
\delta \omega^{ij}_a
&=& \UTT g_{a c} \OT T^{b[i} D_{b}\delta \OT T ^{j] c} 
+ \OT T^{b [i} \UT t^{j]}_c \UT t^k_a D_{b} 
\delta \OT T^c_{k}
\\
&&+ \UT t^{[i}_b D_{a}\delta \OT T^{j] b} 
+ \UT t^k_c \UT t^{[i}_a \OT T^{j]b} D_{b} 
\delta \OT T^c_{k} \ .
\end{eqnarray*}
Then an integration by parts in the varied Hamiltonian yields 
\begin{eqnarray*}
\dot{K}^i_a
&=& - \epsilon^{ijk} K^j_a \Omega^k_{0}
+ N^b D_b K^i_a + D_a N^b K^i_b
\\
&&- \UT N \OT T^b_j
({}^{3}\!R^{ij}_{ab} + K^i_a K^j_b - K^i_b K^j_a) \\
&& + \OT T^b_i D_a D_b \UT N \ .
\end{eqnarray*}

\section{Projectability of gauge transformations} 
\label{sec:project}

We first consider triad gauge rotations.  Under an infinitesimal
rotation with descriptor $\lambda^{i}$ the resulting variation of the
triad field is
\begin{equation}
\delta_{R}[\lambda] \OT T^a_i
= -\epsilon^{ijk} \lambda^j \OT T^a_k \ , \end{equation}
while the variation of $K^{i}_{a}$ is
\begin{equation}
\delta_{R}[\lambda] K^{i}_{a} = -\epsilon^{ijk} \lambda^j K^{k}_{a}\ . 
\end{equation} The Ricci rotation coefficients transform as a
connection in the usual manner under this rotation: \begin{equation}
\delta_{R}[\lambda] \Omega^{i}_{0} = -\partial_{0} \lambda^{i}
-\epsilon^{ijk} \lambda^{j}\Omega^{k}_{0}\ .  \label{rot}
\end{equation}

The Lagrangian density does not depend on time derivatives of $N$,
$N^{a}$, or $\Omega^{i}_{0}$.  Thus, null vectors of the Legendre
matrix (the second partial derivative of the Lagrangian density with
respect to the velocities) are $\partial/\partial\dot N$,
$\partial/\partial\dot N^{a}$, and
$\partial/\partial\dot\Omega^{i}_{0}$; in other words, projectable
symmetry variations under the Legendre transformation may not depend
on $\dot N$, $\dot N^{a}$, or $\dot\Omega^{i}_{0}$, as discussed in
\cite{pons/salisbury/shepley/97}.  As in the conventional formulation
of general relativity, the variations of lapse and shift are not
projectable unless the descriptors $\epsilon^{\mu}$ of an
infinitesimal coordinate transformation $x^{\mu} \rightarrow x^{\mu} -
\epsilon^{\mu}$ depend on the lapse and shift in the following manner:
\begin{equation} \epsilon^{\mu} = \delta^{\mu}_{a} \xi^{a} + n^{\mu}
\xi^{0} \ ,
\label{3.1}
\end{equation}
where $n^{\mu}= (N^{-1}, -N^{-1} N^{a})$ is the normal to the constant
coordinate time hypersurface, the $\xi^{\mu}$ being arbitrary
functions.

We must now check whether variations of $\Omega^{ij}_{0}$ are
projectable under these diffeomorphisms.  We represent the variations
resulting from infinitesimal perpendicular diffeomorphisms, with
descriptors $\epsilon^{\mu}=n^{\mu}\xi^{0}$, by $\delta_{PD}$, where
$PD$ denotes perpendicular diffeomorphism.  (Since variations under
diffeomorphisms for which $\delta x^{0} = 0$ are projectable---they
are the usual Lie derivatives of spatial vectors under spatial
diffeomorphisms---we will not consider them here.)  The objective of
this calculation is the evaluation of the variation of
$\Omega^{ij}_{0}$ on-shell, that is, using the equations of motion. 
In this sense, $\Omega^{ij}_{0}$ should not here be considered as an
independent variable.

Let us first calculate the variations of the tetrad vectors.  We find
\begin{eqnarray*} \delta_{PD}[\xi^{0}] E^0_{0} &=& -N^{-2} \dot
\xi^{0}+ N^{-2} N^{a} \xi^{0}_{,a} \ , \\
\delta_{PD}[\xi^{0}] E^0_{i}
&=&0 \ ,
\\
\delta_{PD}[\xi^{0}] E^a_{ 0}
&=& N^{-2} N^{a} \dot \xi^{0} - N^{-2} N^{a} N^{b} \xi^{0}_{,b} 
\\
&&-N^{-1} e^{a b}N_{,b} \xi^{0}+e^{a b} \xi^{0}_{,b} \ , 
\\
\delta_{PD}[\xi^{0}] E^a_{i}
&=& \delta_{PD}[\xi^{0}] T^a_{i}
\\
&=& \dot T^{a}_{i} N^{-1} \xi^{0}
- T^{a}_{i,b} N^{b} N^{-1} \xi^{0}
\\
&& + N^{a}_{,b}N^{-1} T^b_i \xi^{0} \ .
\end{eqnarray*}
The equation of motion (\ref{EoM}) is equivalent to \[
\dot T^{a}_{i} = -\Omega^{ij}_{0} T^{a}_{j} 
- D_{b} N^{a} T^{b}_{i}
- N T^{b}_{i} T^{a}_{j} K^{j}_{b} \ .
\]
The final expression for the last variation may consequently be
rewritten as
\[
\delta_{PD}[\xi^{0}] T^{a}_{i} = -\xi^{0} \Omega^{ij}_{\mu}n^{\mu}
T^{a}_{j} - \xi^{0} T^{b}_{i} T^{a}_{j} K^{j}_{b} \ .  \] Since we
shall require the result below, we also include here the corresponding
variation of the densitized triad: \begin{equation}
\delta_{PD}[\xi^{0}] \OT T^{a}_{i} = -\xi^{0} \Omega^{ij}_{\mu}n^{\mu}
\OT T^{a}_{j} - \xi^{0} T^{b}_{i} \OT T^{a}_{j} K^{j}_{b} + \xi^{0}
T^{a}_{i} \OT T^{b}_{j} K^{j}_{b} \ .
\label{trdvar}
\end{equation}
Similarly, we find that
\begin{eqnarray}
\delta_{PD}[\xi^{0}] e^0_{ 0}
&=& \dot \xi^{0}-N^{a} \xi^{0}_{,a} \ ,
\label{vare.00}\\
\delta_{PD}[\xi^{0}] e^0_{a}&=&0 \ ,
\label{vare.0a}\\
\delta_{PD}[\xi^{0}] e^i_{0}
&=&-N^{a}\Omega^{ij}_{\mu} n^{\mu}\xi^{0} t^{j}_{a} 
+ N^{a}\xi^{0}K^{i}_{a} \nonumber \\
&& -N_{,a} T^{ai} \xi^{0} + T^{ai} \xi^{0}_{,a} \ , 
\label{vare.i0}\\
\delta_{PD}[\xi^{0}] e^i{}_{\, a}
&=& \delta_{PD}[\xi^{0}] t^i{}_{\, a}
= -\xi^{0}\Omega_{\mu}^{ij}n^{\mu}t^{j}_{a}+\xi^{0}K^{i}_{a}\ . 
\label{vare.ia}
\end{eqnarray}

We shall also require the variation of the densitized lapse $\UT N$. 
Referring to (\ref{vare.ia}) to compute $\delta_{PD}[\xi^{0}]t$, we
find
\begin{eqnarray*}
\delta_{PD}[\xi^{0}] \UT N
= -t^{-1} N \xi^{0} K^{i}_{a} T^{a}_{i} + t^{-1} \dot \xi^{0} 
- t^{-1}N^{a} \xi^{0}_{,a} \ .
\end{eqnarray*}
Using
\[
\dot t = t D_{a}N^{a} + t N T^{a}_{i} K^{i}_{a} \ ,\ 
{}^{3}\Gamma^{b}_{ba} = t^{-1} t_{,a} \ , \]
we find
\begin{equation}
\delta_{PD}[\xi^{0}]\UT N
= \UT{\dot\xi^{0}} + \UT \xi^{0} N^{a}_{,a} 
- N^{a} \UT \xi^{0}_{,a} \ .
\label{dPDN}
\end{equation}

Substituting the variations of the tetrad vectors into (\ref{ricrot}),
we find
\begin{eqnarray}
\delta_{PD}[\xi^{0}] \Omega_{0}^{ij}
&=& 2 N N^{a} K^{[i}_{a}T^{j]b}(N^{-1} \xi^{0})_{,b} 
\nonumber\\
&&+2 N N_{,a}T^{a[i} T^{j]b} (N^{-1}\xi^{0})_{,b} 
\nonumber\\
&&+(\Omega_{\mu}^{ij} n^{\mu} \xi^{0})_{,0} 
+\dot\Omega_{a}^{ij} N^{-1}N^{a} \xi^{0}
\nonumber\\
&&-\Omega_{0,a}^{ij} N^{-1}N^{a} \xi^{0} \ . 
\label{deltaOmegaa}
\end{eqnarray}
It is noteworthy that the first two terms in this expression appear
due to the fact that our tetrad vectors are not true four-vectors
(since they are tied to the time foliation).  Otherwise this variation
agrees with the standard result in Einstein-Yang-Mills theory with an
SO(3) connection\cite{pons/salisbury/shepley/99a}.  To proceed further
we must substitute the time derivative
\begin{eqnarray}
\dot \Omega_{a}^{ij}
&=& D_{a} \Omega_{0}^{ij} +2 K_{a}^{[i} T^{j]b} N_{,b} 
\nonumber \\
&&+ 4 N T^{b[i} D_{[a} K^{j]}_{b]}
+ {}^{3}\!R_{ba}^{ij}N^{b} \ .
\label{Omegadot}
\end{eqnarray}
We find ultimately that
\begin{eqnarray}
\delta_{PD}[\xi^{0}] \Omega_{0}^{ij}
&=& -4 \xi^{0} N^{a} D_{[a} K_{b]}^{[i} T^{j]b} 
\nonumber \\
&&+ 2 N^{b} \xi^{0}_{,a} K^{[i}_{b} T^{j]a} +2 N_{,b} \xi^{0}_{,a} 
T^{b[i}T^{j]a}
\nonumber \\
&&+(\Omega_{\mu}^{ij} n^{\mu} \xi^{0})_{,0} 
+2 \xi^{0} n^{\mu} \Omega_{\mu}^{[i} \Omega_{0}^{j]} \ . 
\label{deltaOmegab}
\end{eqnarray}

We have displayed this result in a form, (\ref{deltaOmegab}), in which
it is manifest that the variation is not projectable: The next to the
last term contains time derivatives of all three gauge variables; but
the resolution is manifest as well.  Referring to (\ref{rot}) we see
that the last two terms are precisely a gauge rotation of the SO(3)
connection components $\Omega_{0}^{ij}$ with a descriptor $
\lambda^{i} = - \Omega_{\mu}^{i} n^{\mu} \xi^{0}$.  To obtain a
projectable variation we must accompany the perpendicular
diffeomorphism with an SO(3) gauge rotation with minus this
descriptor.

This is exactly the form of the gauge transformation which must be
added to diffeomorphisms in the Einstein-Yang-Mills case to produce a
projectable variation under the Legendre transformation, as shown in
\cite{pons/salisbury/shepley/99a}.  The demonstration that this is the
required addition here was complicated somewhat by the fact that the
connection is not a one-form under diffeomorphisms which alter the
time foliation.  As mentioned above this is due to the fact that the
tetrad vectors in our $3+1$ decomposition are not four-vectors.

We close this section with the variation of $\Omega_{0}^{i}$ under
spatial diffeomorphisms, with descriptor
$\epsilon^{\mu}=\delta^{\mu}_{a}\xi^{a}$.  We will represent the
variation by $\delta_{SD}[\vec \xi]$ where SD denotes spatial
diffeomorphism.  This variation is indeed the Lie derivative since the
tetrads do transform as manifest four-vectors under infinitesmal
transformations which do not alter the spacetime foliation:
\begin{equation} \delta_{SD}[\vec \xi] \Omega_{0}^{i} =
\Omega_{0,a}^{i} \xi^{a} + \Omega_{a}^{i} \xi^{a}_{,0} \ .
		\label{deltaSDOmega}
\end{equation}
We wish to obtain the on-shell variation which we will later on
compare with variations generated by symmetry generators which we
construct below.  For this purpose it is convenient to rewrite
(\ref{deltaSDOmega}) as
\begin{eqnarray}
&&\delta_{SD}[\vec \xi] \Omega_{0}^{i}
=(D_{a}\Omega_{0}^{i} - \dot\Omega_{a}^{i})\xi^{a}
+(\Omega^{i}_{a}\xi^{a})_{,0} 
- \Omega^{ij}_{a}\xi^{a} \Omega^{j}_{0}
\nonumber \\
\quad&&\quad= -\epsilon^{ijk}\xi^{a} (K_{a}^{j}T^{bk} N_{,b} 
+ 2 N T^{bj} D_{[a} K_{b]}^{k}) 
\nonumber \\
\quad&&\qquad- {}^{3}\!R_{ba}^{i} N^{b} \xi^{a} 
+ (\xi^{a}\omega_{a}^{i})_{,0}
+\epsilon^{ijk}\xi^{a}\omega_{a}^{j}
\Omega_{0}^{k}\ .
\label{deltaOmegac}
\end{eqnarray}
where in the second equality we used the time derivative
(\ref{Omegadot}).

\section{Generators, variations, and Lie algebra} 
\label{sec:generators}

Our next task is to construct the generators of gauge transformations
in order to verify that the phase space calculations reproduce the
above configuration-velocity space results.  For this purpose we first
introduce generators associated with our secondary constraints:
\begin{eqnarray*} R[\xi] &:=& \int d^3 x \,\xi^i \OT {\cal H}_{ i} \ ,
\\ V[{\vec \xi}] &:=& \int d^3 x \,\xi^a \OT{\cal H}_{ a} \ , \\
S[\UT\xi^{0}] &:=& \int d^3 x \,\UT\xi^0 \OTT {\cal H}_{ 0} \ . 
\end{eqnarray*} We find that $R[\xi]$ generates an SO(3) gauge
rotation, so we have, for example,
\begin{eqnarray*}
\delta_{R}[\xi] \OT T^a_i &=& \{ \OT T^a_i, -\epsilon_{jkl} 
\int d^3 x' \,\xi'^j K'^k_b \OT T^b_l \} \\ 
&=& -\epsilon_{ijl} \xi^j \OT T^a_l \ .
\end{eqnarray*}

$V[{\vec \xi}]$ generates a spatial diffeomorphism plus SO(3) gauge
rotation:
\begin{eqnarray*}
\delta_{V}[\vec\xi] \OT T^a_i
&=&\{\OT T^a_i,-2\int d^3 x'\xi'^b\OT T'^c_i D'_{[c} K'^i_{b]}\}\\ 
&=& 2 D_b ( \xi^{[b} \OT T^{a]}_i)\\
&=& {\cal L}_{\vec \xi} \OT T^a_i + \delta_{R}[\xi^b \omega_b] \OT
T^a_i \ .  \end{eqnarray*} It is convenient to define a related
generator $D[{\vec \xi}]$ which generates a pure spatial
diffeomorphism: \[ D[{\vec \xi}] := \int d^3 x \,\xi^a \OT{\cal G}_{a}
\ , \] where
\[
\OT{\cal G}_{a}:=\OT{\cal H}_{ a}-\omega_{a}^{i}\OT{\cal H}_{i}\ . \]

$S[\UT\xi^{0}]$ generates a space-time diffeomorphism, with descriptor
$\xi^{0}=t\UT\xi^{0}$, plus a gauge rotation (neither of which by
itself is projectable).  So, for example, \begin{eqnarray*} \delta_{S}
[\UT \xi^{0}] \OT T^a_i &=& \delta_{PD}[t \UT \xi^{0}] \OT T^{a}_{i} +
\delta_{R}[t \UT \xi^0 \omega_\mu n^\mu] \OT T^{a}_{i} \\
&=& -\UT \xi^{0} \OT T^{b}_{i} \OT T^{a}_{j} K^{j}_{b} 
+ \UT \xi^{0} \OT T^{a}_{i} \OT T^{b}_{j} K^{j}_{b} \ , \end{eqnarray*}
where in the last line we used (\ref{trdvar}). 

We now turn to the calculation of the complete Lie algebra in
configuration-velocity space.  It would be straightforward to
calculate the algebra from the calculable action of the infinitesimal
group elements on the generators.  The only Poisson bracket we will
not calculate in this manner is the bracket of $S[\UT\xi^{0}]$ with
$S[\UT\eta^{0}]$, simply because it would be tedious, invoking time
derivatives of the triad vectors, the curvature, and the extrinsic
curvature.

First, a gauge rotation of $\OT{\cal H}_{i}$ yields \[
\{R[\xi],R[\eta]\}= -R[[\xi,\eta]] \ ,
\]
where we define the commutator bracket as \[
[\xi,\eta]^{i}:=\epsilon^{ijk}\xi^{j}\eta^{k}\ .  \] In the following
expressions, we will also use the Lie bracket \[
[\vec\xi,\vec\eta]^{a} := \xi^{b}\eta^{a}_{,b} - \eta^{b}\xi^{a}_{,b}
\ .  \] The remaining brackets are
\begin{eqnarray*}
\{R[\xi],D[{\vec \eta}]\}
&=& \int d^{3}x\,\xi^i{\cal L}_{\vec \eta}\OT{\cal H}_{i} \\ 
&=&-\int d^{3}x\,{\cal L}_{\vec \eta}\xi^i\OT{\cal H}_{i} \\ 
&=&-R[{\cal L}_{\vec \eta}\xi] \ ,
\end{eqnarray*}
\begin{eqnarray*}
\{D[{\vec \xi}],D[{\vec \eta}]\}
&=& \int d^{3}x\,\xi^a{\cal L}_{\vec \eta}\OT{\cal G}_{a} \\ 
&=&- \int d^{3}x\,{\cal L}_{\vec \eta}\xi^a\OT{\cal G}_{a} \\ 
&=& -D[{\cal L}_{\vec \eta}{\vec \xi}]
= D[[{\vec \xi},{\vec \eta}]] \ ,
\end{eqnarray*}
\begin{eqnarray*}
\{S[{\UT \xi^{0}}],D[{\vec \eta}]\}
&=&\int d^{3}x\,\UT\xi^0{\cal L}_{\vec \eta}\OTT{\cal H}_{0} \\ 
&=&-\int d^{3}x\,{\cal L}_{\vec \eta}\UT\xi^0\OTT{\cal H}_{0} \\ 
&=& - S[{\cal L}_{\vec \eta}\UT\xi] \ ,
\end{eqnarray*}
\[
\{S[{\UT \xi^{0}}],R[{\eta}]\} = 0 \ ,
\]
\[
\{V[{\vec \xi}],R[{\eta}]\} = 0 \ .
\]
The last two brackets result from the fact that $\OTT{\cal H}_0$ and
$\OT{\cal H}_a$ are gauge rotation scalars.  Finally, direct
calculation yields
\[
\{S[{\UT \xi^{0}}],S[{\UT \eta^{0}}]\}
= V[\vec \zeta]- R[[\xi_{,a}T^{a},\eta_{,b} T^{b}]] \ , \]
where
\[
\zeta^a := (\UT\xi\partial_{b}\UT\eta
- \UT\eta\partial_{b}\UT\xi) \OTT e^{ab} \ . \]

From these brackets we next determine the brackets among the $R$,
$V$, and $S$ generators alone.  We find
\begin{eqnarray*}
\{V[{\vec \xi}],V[{\vec \eta}]\}
&=& \{D[{\vec \xi}]+R[\xi^a \omega_a],D[{\vec \eta}] 
+R[\eta^b \omega_b]\} \\
&=&V[[{\vec \xi},{\vec \eta}]]
-R[{}^{3}\!R_{ab}\xi^a \eta^b] \ .
\end{eqnarray*}
The remaining bracket is
\begin{eqnarray*}
&&\{S[{\UT \xi^{0}}],V[{\vec \eta}]\}
= \{S[{\UT\xi^{0}}],D[{\vec \eta}]+R[\xi^a \omega_a]\} \\ 
&&\qquad= -S[{\cal L}_{\vec \eta}\UT \xi^{0}]
- R[\eta^{a} (\delta_{PD}[t \UT \xi^{0}] 
+\delta_{R}[\Omega_{\mu}n^{\mu}t\UT\xi^{0}]) 
\omega_{a}] \ ,
\end{eqnarray*}
where
\begin{eqnarray*}
(\delta_{PD}[t \UT \xi^{0}]
&+&\delta_{R}[\Omega_{\mu} n^{\mu}t \UT \xi^{0}]) 
\omega_{a}^{i}
\\
&=&(K^{j}_{a} \UT T^{bk}D_{b}(\UT \xi^{0}) 
+2 \OT T^{b}_{j} \UT \xi^{0}D_{[a}K^{k}_{b]} ) 
\epsilon^{ijk} \ .
\end{eqnarray*}

\section{Complete symmetry generators}
\label{sec:symgens}

The canonical Hamiltonian in terms of the generators takes the simple
form
\[
H = \int d^{3}x\,N^A {\cal H}_A
=: N^{A}{\cal H}_{A}\ ,
\]
where we define \[ N^A :=\{\UT N, N^a, -\Omega^i_0\}\ ,\ {\cal H}_A
=\{\OTT{\cal H}_0, \OT{\cal H}_a,\OT{\cal H}_i\}\ , \] and where
spatial integrations over corresponding repeated primed indices are
assumed.  It was shown in \cite{pons/salisbury/shepley/97} that the
complete symmetry generators then take the form \begin{eqnarray} G(t)
= \xi^{A} G^{(0)}_{A} + \dot \xi^{A} G^{(1)}_{A}\ .
\label{Gxi}
\end{eqnarray}
The descriptors $\xi^{A}$ are arbitrary functions.  The simplest
choice for the $G^{(1)}_{A}$ are the primary constraints $P_{A}$, \[
P_{A} := \{\OTT P,\OT P_{a},-\OT P_{i}\}\ , \] with the result that
\[
G[\xi] = P_{A} \dot \xi^{A} + ( {\cal H}_{A} 
+ P_{C''} N^{B'} {\cal C}^{C''}_{AB'}) \xi^{A}\ , \]
where the structure functions are:
\begin{equation}
\{ {\cal H}_{A},{\cal H}_{B'} \}
	=: {\cal C}^{C''}_{AB'} {\cal H}_{C''}\ . \end{equation}

Using the brackets calculated in the previous section we read off the
following non-vanishing structure functions: 
From these brackets we next determine the brackets among the $R$,
$V$, and $S$ generators alone.  We find
\begin{eqnarray*} 
    C^a_{0'0''} 
    &&=\OTT e^{ab}\Big(-\delta^3(x-x')\partial''_b \delta^3 (x-x'')
        \\
    &&\qquad + \delta^3 (x-x'') \partial'_b\delta^3 (x-x') \Big)\ , 
        \\
    C^a_{b' c''}
    &&=-\delta^3 (x-x') \partial''_b\delta^3 (x-x'') \delta^a_c \\ 
    &&\qquad+ \delta^3 (x-x'') \partial'_c \delta^3 (x-x') \delta^a_b\ , \\
    C^0_{0' a''}
    &&=\delta^3 (x-x'') \partial'_a\delta^3 (x-x') \\ 
    &&\qquad-\delta^3 (x-x') \partial''_a\delta^3 (x-x'') \ , \\ 
    C^i_{j' k''}
    &&=-\epsilon^{ijk} \delta^3 (x-x') \delta^3 (x-x'') \ ,\\ 
    C^i_{0' a''}
    &&= \epsilon^{ijk}\Big( K^{j}_{a} T^{bk} t' 
        \partial'_{b}\delta^3 (x-x') \delta^3 (x-x'') \\ 
    &&\qquad -2D_{[a}K^{k}_{b]} \OT T^{b}_{j}\delta^3 (x-x') 
        \delta^3 (x-x'')\Big) \ , \\
    C^i_{a' b''}
    &&= -{}^{3}\!R^i_{ab} \delta^3 (x-x') \delta^3 (x-x'') \ , \\ 
    C^i_{0' 0''}
    && \\
    =&&
    -\epsilon^{ijk}t' t''\Bigl( \partial'_{a} 
    \partial''_{b}\bigl(T'^{aj}T''^{bk}
    \delta^3 (x-x') \delta^3 (x-x'')\bigr)\Bigr) \ . 
\end{eqnarray*}

With the use of the structure functions derived above, we obtain the
following generators, denoted by $G[\xi]$, $G[{\vec\eta}]$, and
$G[\UT\zeta^{0}]$.  These are, respectively, the gauge, spatial
diffeomorphisms, and perpendicular diffeomorphisms (plus associated
gauge rotations in the last two cases):
\begin{eqnarray*}
G[\xi]
&=& \int d^{3}x\, (-\OT P_i \dot{\xi}^{i} 
+\OT{\cal H}_i \xi^{i} - \epsilon_{ijk}\xi^j 
\Omega^k_0 \OT P_i ) \ , \\
G[{\vec \eta}]
&=& \int d^{3}x\, \Big(\OT P_a \dot{\eta}^{a} 
+{}^{3}\!R^i_{ab} \OT P_i \eta^a N^b \\
&&\qquad -\epsilon^{ijk}((t \UT N)_{,b} \eta^{a} 
K^{j}_{a} \OT T^{bk}\OT P_i \\
&&\qquad + 2 D_{[a}K^{k}_{b]} \OT T^{b}_{j} 
\UT N \eta^{a} \OT P_{i})
- \UT N \OTT P \eta^a_{,a}
+ \UT N_{,a} \OTT P \eta^a \\
&&\qquad +N^a_{,b}\OT P_a \eta^b - N^b \eta^a_{,b} 
\OT P_a+\eta^{a}\OT{\cal H}_{a}\Big) \ , \\ G[\UT \zeta^{0}]
&=& \int d^{3}x\, \Bigr(\OTT P \UT {\dot\zeta}^0 
+\UT N_{,b} \OT P_a \UT \zeta^{0} \OTT e^{ab} 
+\UT{\zeta}^{0}\OTT{\cal H}_{0}\\
&&\qquad -\UT N \OT P_a \UT \zeta_{,b}^{0}\OTT e^{ab} 
- N^a \OTT P \UT \zeta_{,a}^{0}
+N^a_{,a} \OTT P \UT \zeta^{0} \\
&&\qquad +\Big((t \UT \zeta^{0})_{,a} (t \UT N)_{,b}\OT T^{aj}\OT
T^{bk} +(t \UT \zeta^{0})_{,b} K^{j}_{a} T^{bk} N^{a} \\
&&\qquad\qquad +2 \UT\zeta^{0}D_{[a}K^{k}_{b]}\OT T^{b}_{j}N^{a}\Big)
\epsilon^{ijk}\OT P_{i} \Bigr) \ .
\end{eqnarray*}
These generators do indeed generate all of the correct symmetry
variations, including the variations of the gauge variables $\UT N$
and $\Omega_{0}^{i}$ displayed in (\ref{dPDN}), (\ref{deltaOmegab}),
and (\ref{deltaOmegac}).

\section{Conclusions}
\label{sec:conclusions}

We have constructed explicitly the full set of generators of
projectable triad rotations, spatial diffeomorphisms, and
perpendicular diffeomorphisms in vacuum General Relativity in a real
triad formalism.  The perpendicular diffeomorphisms are not by
themselves projectable (generated by functions which can be pulled
back to the original configuration-velocity space).  They must be
accompanied by triad rotations which themselves depend on metric,
triad, and extrinsic curvature.  The generators act in a phase space
which includes as variables the gauge functions $N$, $N^{a}$, and
$\Omega^{i}_{0}$.  The group algebra can be deduced from the action of
the transformation group in configuration-velocity space.  We did so
in this paper, except for $\{S[{\UT\xi^{0}}],S[{\UT\eta^{0}}]\}$,
simply because direct calculation of the Poisson bracket was more
efficient in this case.  The manifold on which this enlarged symmetry
group acts must be interpreted to be the set of all solution
trajectories, and the group elements contain a compulsory dependence
on these solutions.  (The pullback from phase space to
configuration-velocity space of variations of arbitrary trajectories
does not generally yield Noether symmetries.  However, the pullbacks
of configuration variables alone do produce Noether symmetries.  See
\cite{pons/salisbury/shepley/99a} and \cite{gracia/pons/99} for
further details.)

Our results for the symmetry variations and corresponding generators
differ technically somewhat from the results in Einstein-Yang-Mills
theories, found in\cite{pons/salisbury/shepley/99a}.  In this paper a
$3+1$ decomposition of the Ricci rotation coefficients destroys the
manifest covariance of these coefficients.  This lack of manifest
covariance is to be contrasted with the Yang-Mills case in which the
connection is a four-dimensional one-form.  The result is a commutator
algebra which differs from the Einstein-Yang-Mills algebra.

Our analysis was undertaken in part to compare and contrast these
results with the symmetry properties of the real sector of Ashtekar's
formulation of general
relativity\cite{ashtekar/86,ashtekar/87,ashtekar/91}.  In fact, many
of the computations performed in this paper are considerably
simplified by working with the canonically transformed complex
variables in Ashtekar's approach.  It does turn out, of course, that
the underlying dynamics in the Ashtekar theory, when restricted to
real triads, does coincide with the real triad dynamics presented in
this paper.  Indeed, the symmetry variations in the complex Ashtekar
formalism do preserve the reality of real triads, and they coincide
with the symmetries presented here.  It does turn out, however, that
the functional form of the gauge rotation required to obtain
projectability of perpendicular diffeomorphisms plus gauge is altered. 
An additional complex triad rotation is required.  These results will
be presented in a forthcoming paper\cite{pons/salisbury/shepley/99b}. 
The issue of projectability is intimately related to the problem of
preservation of reality conditions under time evolution.

The results we have obtained in this and preceding papers represent a
significant conceptual and technical advancement.  In the context of
the Dirac-Bergmann constraint formalism, the conventional wisdom is
that first class phase space generators are to be interpreted as
generators of gauge transformations \cite{dirac/62,dirac/50,dirac/64}. 
Yet this notion has been rejected historically due to the presence of
dynamical fields in the Poisson bracket algebra, leading to what some
call an open algebra; see, for example, \cite{ashtekar/91}.  In these
papers we require that Lagrangian Noether symmetry variations be
projectable to variations in phase space.  We find---amazingly---that
the previously known Hamiltonian constraints generate these projected
phase space gauge variations.  The full four-dimensional
diffeomorphism group is only implementable on shell, that is, on
solutions of the equations of motion.  We conclude that the gauge
group present in configuration-velocity space is projectable to phase
space.  The original full Lagrangian symmetry is therefore retained in
the Hamiltonian formalism.


\section*{Acknowledgments}

JMP and DCS would like to thank the Center for Relativity of The
University of Texas at Austin for its hospitality.  JMP acknowledges
support by CICYT, AEN98-0431, and CIRIT, GC 1998SGR, and wishes to
thank the Comissionat per a Universitats i Recerca de la Generalitat
de Catalunya for a grant.  DCS acknowledges support by National
Science Foundation Grant PHY94-13063.




\end{multicols}

\end{document}